\documentclass[10pt,conference]{IEEEtran}

\usepackage[T1]{fontenc}
\usepackage[utf8]{inputenc}
\usepackage{booktabs}
\usepackage{array}
\usepackage{xspace}
\usepackage{listings}
\usepackage{graphicx}
\usepackage{xcolor}
\usepackage{amsmath}
\usepackage{enumitem}
\usepackage{algorithm}
\usepackage{algpseudocode}
\usepackage{cite}
\usepackage{textcomp}
\usepackage{url}
\usepackage{hyperref}
\usepackage[most]{tcolorbox}

\newcommand{\toolname}{Prefactory\xspace}
\newcommand{\benchmark}{PrefactoryBench\xspace}

\definecolor{pykeyword}{HTML}{0033B3}
\definecolor{pystring}{HTML}{067D17}
\definecolor{pycomment}{HTML}{8C8C8C}
\definecolor{pybuiltin}{HTML}{8959A8}
\definecolor{diffrem}{HTML}{C13030}
\definecolor{diffadd}{HTML}{1E7A2A}

\lstdefinestyle{pythonpretty}{
  language=Python,
  basicstyle=\ttfamily\bfseries\scriptsize,
  keywordstyle=\color{pykeyword},
  numbers=none,
  stringstyle=\color{pystring},
  commentstyle=\color{pycomment}\itshape,
  emph={[2]np,scipy,rankdata,argsort,searchsorted,array,round,astype},
  emphstyle={[2]\color{pybuiltin}},
  showstringspaces=false,
  breaklines=true,
  columns=fullflexible,
  keepspaces=true,
  aboveskip=0pt,
  belowskip=0pt,
  xleftmargin=2pt,
}

\lstdefinestyle{astshape}{
  basicstyle=\ttfamily\scriptsize,
  numbers=none,
  showstringspaces=false,
  breaklines=true,
  columns=fullflexible,
  keepspaces=true,
  aboveskip=0pt,
  belowskip=0pt,
  xleftmargin=2pt,
  emph={[1]Assign,For,AugAssign,Return,BinOp,Add,Div,Num,Call,Name,Subscript},
  emphstyle={[1]\color{pykeyword}\bfseries},
  emph={[2]sum,cnt,idx,val},
  emphstyle={[2]\color{diffadd}\bfseries},
}

\lstdefinestyle{trigger}{
  basicstyle=\ttfamily\scriptsize,
  numbers=none,
  showstringspaces=false,
  breaklines=true,
  breakatwhitespace=false,
  columns=fullflexible,
  keepspaces=true,
  aboveskip=0pt,
  belowskip=0pt,
  xleftmargin=2pt,
  morecomment=[l]{\#},
  commentstyle=\color{pycomment}\itshape,
  emph={[1]name,pattern,weight,target,rationale},
  emphstyle={[1]\color{pykeyword}\bfseries},
}

\newtcolorbox{delband}{
  enhanced jigsaw,
  colback=red!11,
  colframe=red!11,
  boxrule=0pt,
  arc=0pt,
  left=0pt,
  right=0pt,
  top=-1pt,
  bottom=-2pt,
  boxsep=0pt,
  before skip=-2pt,
  after skip=-2pt,
}

\newtcolorbox{addband}{
  enhanced jigsaw,
  colback=green!13,
  colframe=green!13,
  boxrule=0pt,
  arc=0pt,
  left=0pt,
  right=0pt,
  top=-1pt,
  bottom=-2pt,
  boxsep=0pt,
  before skip=-2pt,
  after skip=-2pt,
}

\newtcolorbox{answerbox}{
  enhanced,
  colback=black!6,
  colframe=black!30,
  boxrule=0.4pt,
  arc=2pt,
  left=4pt,
  right=4pt,
  top=3pt,
  bottom=3pt,
  before skip=5pt,
  after skip=5pt,
}

\newtcolorbox{beforebox}[1][]{
  enhanced,
  colback=white,
  colframe=red!55!black,
  fonttitle=\bfseries\scriptsize,
  coltitle=white,
  colbacktitle=red!55!black,
  boxrule=0.4pt,
  arc=2pt,
  left=2pt,
  right=2pt,
  top=0pt,
  bottom=0pt,
  boxsep=1pt,
  before skip=1pt,
  after skip=1pt,
  #1
}

\newtcolorbox{afterbox}[1][]{
  enhanced,
  colback=white,
  colframe=green!45!black,
  fonttitle=\bfseries\scriptsize,
  coltitle=white,
  colbacktitle=green!45!black,
  boxrule=0.4pt,
  arc=2pt,
  left=2pt,
  right=2pt,
  top=0pt,
  bottom=0pt,
  boxsep=1pt,
  before skip=1pt,
  after skip=1pt,
  #1
}

\newtcolorbox{shapebox}[1][]{
  enhanced,
  colback=blue!4,
  colframe=blue!45!black,
  fonttitle=\bfseries\scriptsize,
  coltitle=white,
  colbacktitle=blue!45!black,
  boxrule=0.4pt,
  arc=2pt,
  left=2pt,
  right=2pt,
  top=0pt,
  bottom=0pt,
  boxsep=1pt,
  before skip=1pt,
  after skip=1pt,
  #1
}

\newtcolorbox{triggerbox}[1][]{
  enhanced,
  colback=violet!3,
  colframe=violet!50!black,
  fonttitle=\bfseries\scriptsize,
  coltitle=white,
  colbacktitle=violet!50!black,
  boxrule=0.4pt,
  arc=2pt,
  left=2pt,
  right=2pt,
  top=0pt,
  bottom=0pt,
  boxsep=1pt,
  before skip=1pt,
  after skip=1pt,
  #1
}

\title{\toolname{}: Automated Discovery and Application of Library-Adoption Refactorings}

\author{
\IEEEauthorblockN{Islem Bouzenia}
\IEEEauthorblockA{CISPA Helmholtz Center for Information Security\\
Germany}
\and
\IEEEauthorblockN{Michael Pradel}
\IEEEauthorblockA{CISPA Helmholtz Center for Information Security\\
Germany}
}

\begin{document}
\maketitle

\begin{abstract}
Replacing hand-written code with library API calls is a common refactoring that can reduce code size, make code more idiomatic, and reuse well-tested implementations. Yet many library-adoption opportunities are hard to find automatically: the original code often does not mention the target library and may resemble the library API only in behavior, with little syntactic overlap. Existing tools, such as linters and static modernizers, cover only a small set of manually specified patterns. LLMs and LLM-based agents, on the other hand, can generalize to more patterns, but they are costly, difficult to reproduce and to apply systematically at scale.

This paper introduces \toolname{}, an automated approach for library-adoption refactoring in Python. The key idea is to use an LLM to synthesize executable search heuristics rather than relying on repeated LLM prompting over a codebase. Given a target project and a target library name, \toolname{} collects library metadata and project vocabulary, then generates lexical and structural detectors. \toolname{} executes the detectors during a scan phase to find candidate functions. It then heuristically ranks the candidate functions, generates refactorings for the highest-ranked ones using an LLM, and validates the results with project tests and newly generated differential tests.

We evaluate \toolname{} on \benchmark{}, a benchmark of 100 real-world library-adoption refactorings from 61 open-source Python projects and 18 libraries. \toolname{} detects 75 instances at the file level and 56 at the function level, compared with 35 and 32 for the strongest baseline (Codex CLI). From the 56 detected functions, \toolname{} produces 40 test-validated refactorings. Generated detectors make scanning inexpensive: scanning a project takes 1.3 seconds on average, requires no LLM calls, and keeps end-to-end LLM cost below \$0.05 per instance. These results show that detector synthesis can turn LLM semantic knowledge into a practical, scalable workflow for library-adoption refactoring.

\end{abstract}

\section{Introduction}
\label{sec:introduction}

\begin{figure}
\centering

\begin{beforebox}[title={BEFORE \textnormal{(\texttt{mage\_ai/.../statistics.py}, source commit)}}]
\begin{lstlisting}[style=pythonpretty]
import math

def build_histogram_data(col1, series, column_type):
    ...
    if bucket_interval == 0:
        return
\end{lstlisting}
\begin{delband}
\begin{lstlisting}[style=pythonpretty]
    for value in series.values:
        index = math.floor((value - min_value) / bucket_interval)
        if index >= len(buckets):
            index = len(buckets) - 1
        buckets[index]['values'].append(value)
\end{lstlisting}
\end{delband}
\begin{lstlisting}[style=pythonpretty]
    ...
\end{lstlisting}
\end{beforebox}

\vspace{2pt}

\begin{afterbox}[title={AFTER \textnormal{(same file, refactoring commit)}}]
\begin{lstlisting}[style=pythonpretty]
import math
\end{lstlisting}
\begin{addband}
\begin{lstlisting}[style=pythonpretty]
import numpy as np
\end{lstlisting}
\end{addband}
\begin{lstlisting}[style=pythonpretty]

def build_histogram_data(col1, series, column_type):
    max_value = series.max()
    min_value = series.min()
    buckets, bucket_interval = build_buckets(
        min_value, max_value, BUCKETS, column_type)
    if bucket_interval == 0:
        return
\end{lstlisting}
\begin{addband}
\begin{lstlisting}[style=pythonpretty]
    counts, edges = np.histogram(
        series.values, bins=BUCKETS,
        range=(min_value, max_value))
\end{lstlisting}
\end{addband}
\begin{lstlisting}[style=pythonpretty]
    ...
\end{lstlisting}
\end{afterbox}

\caption{Motivating library-adoption refactoring from \texttt{mage\_ai}.}
\label{fig:motivating}
\end{figure}

Library-adoption refactoring replaces hand-written implementations with calls to existing library APIs, preserving program behavior. This refactoring is common in Python projects, where developers often reimplement functionality already provided by libraries such as \texttt{numpy}, \texttt{pandas}, \texttt{scipy}, or \texttt{sklearn}. Replacing such code with library calls can reduce code size, make implementations more idiomatic, and reuse code maintained and tested by library developers.

Figure~\ref{fig:motivating} shows a representative example from the \texttt{mage\_ai} project. The original implementation manually builds histogram buckets by iterating over a series, computing a bucket index, and updating the selected bucket. The developer's refactoring replaces the hand-written histogram computation with \texttt{np.histogram}. Although the refactoring changes only a specific region of code, automatically finding such opportunities in a large project is non-trivial. The original code in this example does not mention the target library near the refactored region. Its relation to \texttt{np.histogram} is visible only through implementation-level signals: a loop over values, a bucket-index computation, updates to histogram buckets, and the domain-specific term ``histogram'' in the name of the surrounding function.

Existing tools provide limited support for this kind of refactoring. Linters and static modernizers, such as \texttt{ruff}~\cite{ruff}, \texttt{refurb}~\cite{refurb}, and \texttt{pylint}~\cite{pylint}, are effective when a refactoring opportunity can be encoded as a precise rule, but cover only patterns defined in advance. Prior work on \emph{custom-to-API replacement}~\cite{tufano2022retiwa}, including RETIWA~\cite{tufano2022retiwa} and AKIRA~\cite{nyirongo2026akira}, also replaces hand-written implementations with calls to existing APIs. However, these approaches typically assume predefined transformation patterns or API mappings and have primarily targeted Java. In contrast, library-adoption refactoring in Python requires detecting opportunities for a target library even when the code does not mention the library, the relevant API, or a fixed syntactic pattern. LLMs and LLM agents are more flexible and can recognize more varied implementations, but using them directly to explore entire projects is costly, difficult to control, and hard to reproduce.

These limitations make \emph{detection} the first bottleneck for library-adoption refactoring. An approach to this problem must address two detection challenges. First, it must localize relevant code without relying on explicit library mentions: the hand-written implementation may not import the target library or name the relevant API. Second, it must generalize beyond fixed syntactic templates: libraries span many domains and expose APIs for many kinds of behavior, and developers can implement the same functionality in different local forms. Once a candidate is found, a complementary task remains: the approach must generate a library-based replacement and reject edits that change the behavior of the original implementation.

The histogram example in Figure~\ref{fig:motivating} illustrates the two detection challenges. The original code does not call \texttt{np.histogram}, but it contains lexical cues, such as domain terms including \texttt{histogram}, \texttt{bucket}, and \texttt{bin}. It also contains structural cues, such as a loop over values, arithmetic used to compute a bucket index, and accumulation into a data structure. These cues are not tied to a single syntax: another developer could implement the same histogram computation using different variable names, control flow, or helper functions. At the same time, they recur across implementations of the same underlying behavior, which makes them useful as reusable detection cues. \toolname{} uses an LLM to synthesize detector families that capture such lexical and structural cues while tolerating variation in local implementation details.

To address these challenges, this paper introduces \toolname{}, an automated approach for library-adoption refactoring in Python. The key insight of our approach is to use an LLM to synthesize \emph{reusable executable detectors} rather than repeatedly prompt an LLM to inspect a codebase. These detectors turn the LLM's semantic knowledge about library functionality into deterministic search mechanisms that can be applied efficiently across a project.

\toolname{} is organized as a four-stage pipeline. First, in the \emph{detector-generation} stage, \toolname{} collects target-library metadata and project vocabulary, then uses an LLM to generate two detector families: lexical detectors, implemented as regular expressions, and structural detectors, implemented as AST matchers. Second, in the \emph{project-scanning} stage, \toolname{} executes the generated detectors over the target project without LLM calls, producing candidate files and functions that may contain library-adoption opportunities. Third, in the \emph{candidate-ranking and filtering} stage, \toolname{} prioritizes candidates using signals such as detector-match strength, matched-code size, and the presence of basic Python implementation patterns, such as loops, branches, arithmetic, and updates to data structures. These patterns help identify hand-written code that may implement functionality available through higher-level library APIs. This stage limits LLM-based rewriting to a small set of localized functions. Fourth, in the \emph{refactoring-generation and validation} stage, \toolname{} invokes an LLM to generate a library-based edit for the highest-ranked functions, then validates the edit with LLM-generated differential tests followed by project tests. The differential tests compare the original and refactored implementations on generated inputs, using the original implementation as the oracle for observed behavior.

To evaluate our technique, we created \benchmark{}, a benchmark of 100 real-world library-adoption refactorings mined from 61 open-source Python projects and spanning 18 libraries, including both third-party libraries, such as \texttt{numpy} and \texttt{pandas}, and standard-library targets, such as \texttt{collections} and \texttt{itertools}. Each benchmark instance includes a reference execution environment and a test setup that pairs the \emph{project tests} (the project's own test suite) with LLM-generated \emph{held-out tests} targeting the refactored region. During its final stage, \toolname{} runs only the project tests; held-out tests are never seen by \toolname{} and exist solely for benchmarking. Both are separate from the differential tests that \toolname{} generates internally during its final stage.

On this benchmark, \toolname{} detects 75 instances at the file level and 56 at the function level, outperforming linter-based tools, vanilla LLM prompting, Semgrep+LLM, and Codex CLI. The strongest baseline, Codex CLI, detects 35 instances at the file level and 32 at the function level while using the same underlying model. \toolname{} also produces 40 validated refactorings from the 56 detected functions sent to generation. The approach scans projects in 1.3 seconds on average and costs less than \$0.05 per instance to detect and refactor; Codex CLI costs roughly three times more.

In an open-source study on the latest versions of five benchmark projects, \toolname{} surfaces multiple opportunities per repository. Two of five submitted pull requests have already been accepted by maintainers, providing initial evidence that the pipeline can surface useful refactorings beyond historical benchmark commits.

In summary, this paper makes the following contributions:
\begin{enumerate}[leftmargin=*,nosep]
\item \textbf{\benchmark{}:} a benchmark of 100 real-world Python library-adoption refactorings across 18 libraries and 61 projects, with reference execution environments, project tests, and held-out tests for evaluating both detection and transformation.

\item \textbf{\toolname{}:} an automated approach based on the idea that LLMs can synthesize reusable executable detectors for library functionality. Starting from a target library name, \toolname{} collects library metadata and project vocabulary, synthesizes lexical and structural detectors, and scans projects without LLM calls during detection.

\item \textbf{Evidence that detector synthesis outperforms direct prompting and agentic search:} a controlled comparison against linter-based tools, vanilla LLM prompting, Semgrep+LLM, and Codex CLI on the 100-instance benchmark, with an end-to-end evaluation and an open-source study on current project versions.
\end{enumerate}

\section{Approach}
\label{sec:approach}

\begin{figure*}[t]
\centering
\includegraphics[width=\textwidth]{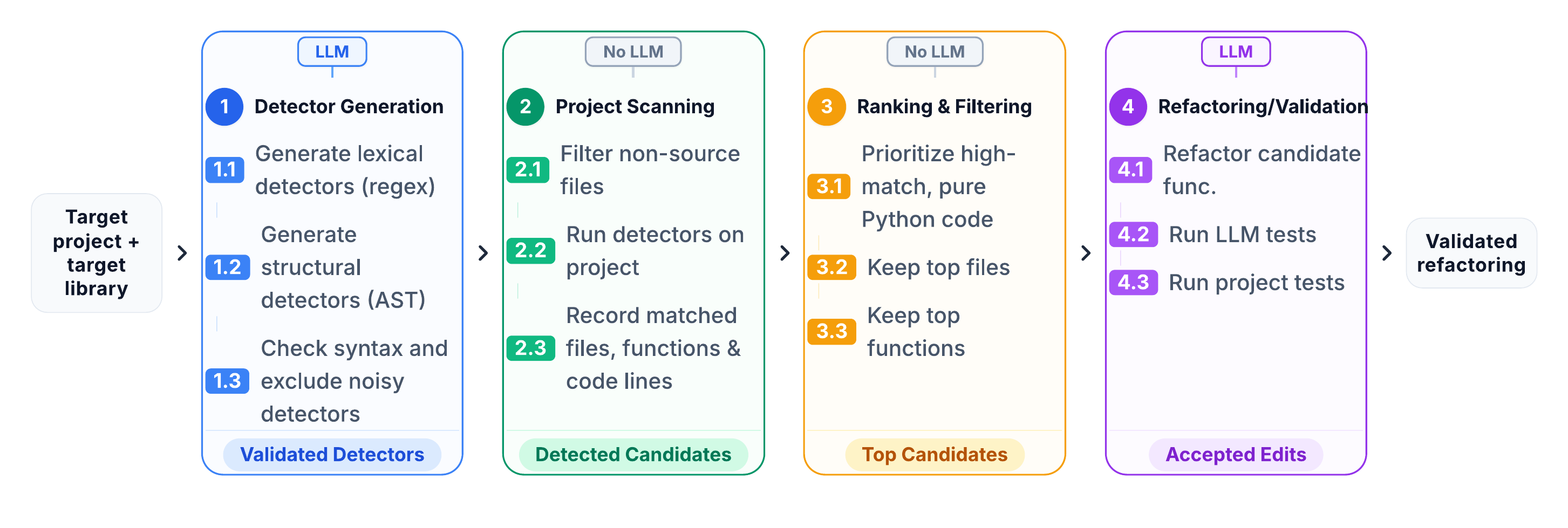}
\vspace{-3em}
\caption{Overview of the \toolname{} four-stage pipeline. Stages~1 and~4 invoke an LLM; Stages~2 and~3 are deterministic and require no LLM calls.}
\label{fig:overview}
\end{figure*}

Figure~\ref{fig:overview} shows \toolname{} as a four-stage narrowing pipeline. Stage~1, \emph{detector generation}, uses an LLM to synthesize lexical and structural detectors and discards detectors that are malformed, fail basic execution checks, or produce overly broad matches. Stage~2, \emph{project scanning}, executes the remaining detectors over the target project, without relying on LLM calls, to find candidate code regions. Stage~3, \emph{candidate ranking and filtering}, ranks the detected files and functions and keeps only a small set of high-ranked candidates. Stage~4, \emph{refactoring generation and validation}, uses an LLM to rewrite the selected functions and validates the resulting edits with differential tests and project tests. This design lets \toolname{} search broadly with deterministic detectors before spending LLM refactoring effort on a much smaller set of localized functions.

\subsection{Stage 1: Detector Generation}
\label{sec:approach:detector-generation}

A detector is an executable search program that looks for hand-written code that may be replaceable by a call to a library API. \toolname{} generates two detector families. \emph{Lexical detectors} are regular expressions that match source-code text, such as API-related names, domain terms, local identifiers, and textual patterns associated with hand-written implementations. \emph{Structural detectors} are AST matchers that match code structure, such as loops, branches, arithmetic expressions, indexed updates, accumulator variables, and exception-handling patterns.

\toolname{} uses separate LLM prompts to generate the two detector families. Both prompts include target-library context: package metadata from PyPI for third-party libraries, importable submodules, exposed functions and classes, usage examples, and mined examples of prior library-adoption refactorings. We obtain usage examples from the target library's documentation by asking a cost-efficient LLM (gpt-5-nano) to extract examples from the package-level documentation.
The mined refactoring examples are used as detector-generation hints.
We collect them by searching GitHub commit histories for commits that introduce an import of the target library using its known import aliases, e.g., \texttt{import numpy as np}, and remove hand-written implementations. These examples are drawn from projects that do not overlap with the benchmark projects. The full detector-generation prompts are included in the supplementary material.

For lexical detectors, \toolname{} additionally provides project vocabulary to the generation prompt. It extracts variable names, function names, class names, and imported names from all Python files in the target project, including tests, selects the 100 most frequent names, and includes up to three sampled source lines per name. This project context helps the LLM produce regular expressions that reflect local names, domain terms, and naming conventions. Structural-detector generation does not use project vocabulary, because AST matchers are intended to capture algorithmic and functionality shapes rather than project-specific names.

Figure~\ref{fig:detectors} shows detector examples for the motivating \texttt{np.histogram} refactoring. The lexical detector searches for histogram-related terms, index computations, and bucket updates. The structural detector abstracts away from exact variable names and instead matches the computation shape: iterating over values, deriving a bin index, optionally adjusting bounds, and updating an indexed container.

Before scanning, \toolname{} removes detectors that are malformed, crash during execution, or produce overly broad matches. Lexical detectors must compile and must not match more than 20\% of project files for that project. Structural detectors are generated together with positive and negative Python examples; \toolname{} parses these examples, runs the generated matcher on their ASTs, and keeps the detector only if it matches all positive examples and none of the negative examples. 
\begin{figure}
\centering

\begin{triggerbox}[title={Lexical detector \textnormal{(regular expression)}}]
\begin{lstlisting}[style=trigger]
target   numpy.histogram
name     histogram_bucket_terms
pattern  (histogram|bucket|bin).{0,120}
         (floor|int|//|round).{0,120}
         (append|\+=|count)

# The .{0,120} terms allow up to 120
# characters between related cues.
\end{lstlisting}
\end{triggerbox}

\vspace{2pt}

\begin{shapebox}[title={Structural detector \textnormal{(Python AST matcher, simplified)}}]
\begin{lstlisting}[style=pythonpretty]
def match(stmts, i, params):
  loop = stmts[i]
  if not isinstance(loop, ast.For):
    return None
  if _for_iter_attr(loop.iter) is None:  # for v in X.values
    return None
  bin_idx = bucket_update = False
  for s in loop.body:
    bin_idx |= _assigns_floor_div(s) # idx = floor(...)
    bucket_update |= _subscript_update(s) # buckets[idx] += v
  if bin_idx and bucket_update:
    return Match("numpy", "np.histogram",
                 span=(loop.lineno, loop.end_lineno))
  return None
\end{lstlisting}
\end{shapebox}

\caption{Detector examples. Top: a lexical detector for \texttt{np.histogram}; Bottom: a simplified AST matcher for \texttt{np.histogram}; each shipped matcher is a per-pattern Python \texttt{match()} function that walks AST nodes and returns the matched span and target API.}
\label{fig:detectors}
\end{figure}

\subsection{Stage 2: Project Scanning}
\label{sec:approach:project-scanning}

Project scanning runs the Stage~1 detectors over the target project at a specific commit.

Before scanning, \toolname{} removes files that are unlikely to contain relevant source code. The approach identifies these files using path, filename, and extension heuristics, e.g., vendored-code directories such as \texttt{\_vendor} or \texttt{\_vendoring}, documentation directories and non-source extensions, virtual-environment paths, generated-file locations, common test directories, and \texttt{test\_*.py} naming patterns. Test files are retained only when the target library is test-related, which \toolname{} determines from PyPI metadata and a fixed list of test-related libraries.

For each remaining file, \toolname{} runs the lexical and structural detectors produced in Stage~1. When a detector matches, \toolname{} records the matched lines, the detector that matched them, the enclosing function, and the containing file. A candidate region may be detected by lexical detectors, structural detectors, or both. The matched lines and matching detectors are retained to guide ranking and to give the generator context on why the candidate was flagged.

In the running example, scanning the \texttt{mage\_ai} project with \texttt{numpy} as the target library identifies the file containing \texttt{build\_histogram\_data}. The generated detectors match histogram-related terms and bucketization-like code structure. The resulting candidate region, together with the detectors that matched it, is passed to the ranking stage.

\subsection{Stage 3: Candidate Ranking and Filtering}
\label{sec:approach:ranking}

Scanning may surface many candidates, not all of them equally strong. \toolname{} therefore ranks them and keeps only the top-ranked candidates before generation.
For each candidate file or function $c$, \toolname{} computes a ranking tuple $(D(c),\, B(c),\, -S(c))$.
The first component, $D(c)$, is detector-match strength:
\[
  D(c) = \bigl(M(c) + U(c)\bigr) \cdot P(c),
\]
where $M(c)$ is the weighted number of detector matches, $U(c)$ is the weighted number of unique matching detectors, and $P(c)$ is the fraction of source lines in $c$ matched by at least one detector. Structural detector matches receive weight~2 and lexical detector matches receive weight~1. We use this fixed 2:1 weighting because structural detectors encode control-flow and data-flow shapes and are therefore more specific than lexical cues; the weights are fixed across all experiments. Including $U(c)$ in the score reduces the effect of size: a large candidate with many matches from a single detector is weaker evidence than one matched by several independent detectors.

The second component, $B(c)$, measures the density of basic Python implementation patterns that often appear in hand-written implementations of higher-level library functionality. These patterns include pure-Python loops, arithmetic operations, branches, updates to simple data structures, and imports from modules that are lower-level with respect to the target library. For a third-party target library, \toolname{} treats the library's declared dependencies and Python standard-library modules as lower-level imports with respect to that library. For example, when ranking candidates for \texttt{numpy}, code that combines arithmetic-heavy loops with imports from standard-library modules such as \texttt{math} receives higher priority.

The final component, $S(c)$, is candidate size, measured as the number of source lines in the file or function. Using $-S(c)$ in the tuple breaks ties by preferring shorter candidates.

Given the ranking tuple, \toolname{} compares candidates lexicographically: it first compares detector-match strength $D(c)$, then the basic-implementation score $B(c)$, and finally candidate size $S(c)$ as a tie breaker.
The approach applies the same ranking rule at two levels: it first keeps the top five files, then ranks functions within those files and keeps the top ten functions overall. This fixed budget bounds LLM generation cost in Stage~4. We study the sensitivity of this budget in Section~\ref{sec:evaluation}.

\subsection{Stage 4: Refactoring Generation and Validation}
\label{sec:approach:generation}

The refactoring-generation and validation stage takes the top-ranked candidates and attempts to rewrite each toward using the target library. Each attempt receives the candidate function, 100 lines of surrounding file context, target-library API hints, and the detector information for the candidate. The API hints come from the same target-library context used in Stage~1, including package metadata, importable submodules, exposed functions and classes, and usage examples. The detector information includes the detector names, target API families, detector patterns, and exact code lines that matched. The result is either a validated refactoring or a rejected candidate.

First, \toolname{} prompts an LLM to rewrite the candidate function using APIs from the target library. The prompt contains three kinds of information: i) the candidate function and local file context, consisting of surrounding code before and after the function; ii) library hints describing the target library and relevant APIs; and iii) the detectors that matched the candidate function and the exact code lines they matched. If the target library exposes at most 100 importable functions and classes, \toolname{} passes the full list to the refactoring prompt. If it exposes more than 100 APIs, \toolname{} first provides the full API list to an LLM and asks it to select up to 100 APIs that are likely relevant to the candidate function, then passes only that selected list to the refactoring prompt. The LLM is then asked to return either a concrete refactoring of the candidate function, together with any necessary imports, or to report that no suitable refactoring is possible.

After the LLM suggests a rewrite, \toolname{} applies a relevance filter. The filter is AST-based: it rejects edits that only add an import without changing the implementation, and edits that do not call the target library in the rewritten function. Candidates that pass this filter are then checked with differential tests and project tests.

To compare the original and rewritten implementations, \toolname{} uses a second LLM call to generate a standalone differential test script~\cite{mckeeman1998differential}. The script places the original function and the rewritten function side by side, includes required dependencies available from the file context when possible, mocks remaining dependencies when needed, generates test inputs, and compares the outputs using the original function as the behavioral oracle, a well-established choice under test-oracle scarcity~\cite{barr2015oracle}. The differential test passes when none of the generated inputs reveal a behavioral difference, fails when at least one produces diverging outputs, and is inconclusive when the script cannot execute reliably, for example because of dependency or mocking failures. The prompt asks the LLM to generate more than ten inputs, including edge cases. A passing differential test is therefore evidence of behavioral preservation on the generated inputs, not a proof of equivalence.

If the differential test passes, \toolname{} runs project tests on the refactored version and accepts the refactoring if no regressions are found. If the differential test fails, \toolname{} feeds the failing inputs and observed outputs back to the refactoring prompt and requests a corrected edit, following the feedback pattern of conversational program repair~\cite{xia2023conversational}. If the differential test is inconclusive because the generated test script errors during setup, mocking, or execution, \toolname{} asks the LLM to repair the test script instead of changing the refactoring edit. \toolname{} allows up to three additional generation attempts. Candidates that still fail relevance filtering, differential testing, or project tests are rejected.

In the running example, this stage rewrites the detected \texttt{build\_histogram\_data} function using \texttt{np.histogram}. \toolname{} validates the rewritten function against the original with differential tests, then accepts the edit after the project tests pass without new regressions.

\section{Evaluation}
\label{sec:evaluation}

In evaluating \toolname{}, we aim to answer five questions:

\begin{description}
\item[RQ1: End-to-End Effectiveness.] How often does \toolname{} detect real-world library-adoption opportunities and produce validated refactorings, and where does it fail?

\item[RQ2: Comparison to Baselines.] How does \toolname{} compare with the baselines on detection and refactoring?

\item[RQ3: Efficiency.] What are the runtime, LLM-call, and monetary costs of \toolname{}?

\item[RQ4: Ablation Study and Design Contributions.] How does \toolname{}'s design affect detection and refactoring?

\item[RQ5: Open-Source Usefulness.] Can \toolname{} surface actionable refactorings on current project versions?

\end{description}

We describe the benchmark, baselines, metrics, and model configuration in turn.

\subsection{Experimental Setup}
\label{sec:eval:setup}

\subsubsection{Benchmark}
\label{sec:eval:benchmark}

Existing refactoring benchmarks do not directly support our task end to end. Some focus on specific classes of Python refactorings, such as replacing non-idiomatic constructs with predefined Pythonic idioms~\cite{zhang2022idiomatic}. Recent repository-level benchmarks such as SWE-Refactor~\cite{xu2026swerefactor} provide high-quality real-world refactoring instances, but target general refactoring scenarios in Java rather than Python library-adoption refactorings. Our setting instead requires instances where a hand-written Python implementation is replaced by a library API call, together with project context to detect the opportunity, generate a replacement, and validate the resulting edit.

To fill this gap, we construct \benchmark{}, a benchmark of 100 real-world library-adoption refactorings mined from 61 open-source Python projects and spanning 18 libraries. The benchmark is designed to facilitate a reliable end-to-end evaluation of refactoring techniques such as \toolname{}: each instance is manually checked, tied to the developer's original refactoring, paired with a reference execution environment, and accompanied by tests that exercise the refactored region. Each instance consists of a parent commit, a developer refactoring commit, the reference file and changed region modified by the developer, and a reference execution environment. The benchmark covers both third-party libraries, such as \texttt{numpy} and \texttt{pandas}, and standard-library targets, such as \texttt{collections} and \texttt{itertools}.

\begin{table}
\centering
\caption{Composition of \benchmark{}.}
\label{tab:benchmark-composition}
\resizebox{\columnwidth}{!}{%
\begin{tabular}{lr}
\toprule
\textbf{Property} & \textbf{Value} \\
\midrule
Library-adoption refactorings & 100 \\
Open-source projects & 61 \\
Target libraries & 18 \\
Median refactored-region size & 30 LOC \\
\midrule
Project-test coverage of refactored region (avg.\ / med.) & 3\% / 4\% \\
Benchmark-suite coverage after held-out tests (avg.) & 94\% \\
Instances with 100\% benchmark-suite coverage & 28 \\
\bottomrule
\end{tabular}%
}
\end{table}

We mine instances from open-source Python repositories on GitHub. We select third-party target libraries by starting from the top 200 PyPI packages by download count, reranking them by in-the-wild import frequency measured through GitHub code search, and keeping libraries with well-documented idioms that replace recognizable hand-written implementations. We supplement these libraries with five widely used standard-library modules. For each target library, we collect active Python repositories using library-specific import queries, such as \texttt{import numpy as np} or calls through common aliases such as \texttt{np.*}. We restrict the search to projects with at least 100 stars, scan commits from 2016 to 2026 using PyDriller~\cite{spadini2018pydriller}, cap each repository at its 500 most recent commits, and skip merge commits.

A diff hunk becomes a candidate when the post-commit file imports the target library and the added lines contain a call to that library API, which we validate using AST-based checks. We deliberately avoid requiring the deleted lines to match any predefined hand-written pattern, so that the benchmark does not favor specific refactoring shapes. We then apply an LLM filter to remove cosmetic substitutions, deprecation swaps between library versions, and changes bundled with broader rewrites that alter functionality or behavior beyond the refactoring. Each remaining instance is manually audited to confirm that it represents a genuine library-adoption refactoring and that the reference file and changed region correspond to the developer's edit.

Finally, we ensure that each benchmark instance is paired with a working test setup. We combine the project tests (i.e., the project's own test suite) with LLM-generated held-out tests that target the refactored region and increase line coverage; together they form the \emph{benchmark test suite}. Held-out tests use the original implementation as the behavioral oracle and pass on both the original and developer-refactored code. During Stage~4 validation, \toolname{} runs only the project tests; held-out tests are never seen by \toolname{} and exist solely to raise test coverage of the target region for benchmark quality assessment. Across the benchmark, the project tests alone cover the refactored region with an average of 3\% and a median of 4\% line coverage, which is insufficient for reliable behavioral validation. Held-out tests targeting the refactored region raise the average benchmark-suite coverage to 94\%, and 28 instances reach 100\% coverage under the full suite.

\subsubsection{Baselines}
\label{sec:eval:baselines}

We compare \toolname{} against four baseline families.

\paragraph{Linters and static modernizers}
We run three Python refactoring and modernization tools: \texttt{ruff}, \texttt{refurb}, and \texttt{pylint}. These tools represent manually authored rule-based detection. We take the union of locations reported by the three tools and calculate file-level and function-level detection rates (see definitions below). This baseline is expected to be precise when a supported rule matches, but it cannot detect opportunities that require rules not implemented by the tools. Linters are evaluated only for detection because they do not provide the same end-to-end refactoring generation workflow as \toolname{}.

\paragraph{Vanilla LLM}
This baseline tests whether a general LLM can localize and refactor library-adoption opportunities without \toolname{}'s generated detectors. The model receives the project file tree and the target library name and is asked to select up to 10 files likely to contain refactoring opportunities, using filenames and paths only. This file budget is larger than \toolname{}'s top-five cutoff because the model has no code-level signal at this step. The model then receives function signatures from the selected files and selects up to 10 candidate functions in total. The code of those functions is then passed again to the LLM to refactor them. This baseline avoids repository-wide code ingestion to keep cost comparable. We validate its suggested refactorings using the same validation setup as for our approach.

\paragraph{Semgrep+LLM}
This baseline uses an LLM to generate rules for the Semgrep~\cite{semgrep} pattern-matching engine, validates the generated rules, and applies them to the project. We evaluate it for detection only. It tests whether a standard pattern-matching engine combined with LLM-written rules can substitute for \toolname{}'s detector generation and ranking.

\paragraph{Codex CLI agent}
We use Codex CLI as an agentic code-editing system to explore a codebase and suggest up to 10 refactorings. We give it the same input as \toolname{} and we ask it to find and refactor candidate opportunities. Codex CLI uses the same underlying model as \toolname{}, allowing us to compare a purpose-built pipeline against an agent using comparable model capability on the same benchmark. 

\subsubsection{Metrics}
\label{sec:eval:metrics}

\paragraph{Detection metrics}
We report detection at two granularities under each approach's fixed reporting budget. A \emph{file-level detection} is correct when the file containing the developer's refactoring appears in the approach's reported candidate files. A \emph{function-level detection} is correct when the function containing the developer-refactored region appears in the approach's reported candidate functions. For \toolname{}, this means the gold file must appear among the top five ranked files, and the gold function must appear among the top ten ranked functions selected from those files. For baselines, we apply the corresponding reporting budget described in Section~\ref{sec:eval:baselines}. Linters and Semgrep+LLM are scored as a boolean per file with no location cap, so firing once or fifty times counts equally; this gives them the most forgiving possible file-level scoring. Additionally, for linters and Semgrep+LLM, function-level detection is scored by mapping each matched line to its enclosing function with no function-count cap applied.

Function-level detection is stricter than file-level detection, but it is more useful for refactoring generation. Passing an entire file to a generator increases LLM cost, makes it harder to create focused behavioral tests, and increases the risk that the model changes irrelevant code. Function-level localization therefore measures whether a tool can narrow the search to a concrete region that is practical to refactor.

\paragraph{Proposed Refactoring and Pass Diff Test}
We count an instance as a \emph{proposed refactoring} (\emph{Refac.}) when the approach detects the correct function and generates a non-trivial edit that passes the relevance filter. We count it as \emph{Pass Diff Test} (\emph{PDT}) when that edit additionally passes the differential tests with up to three repair attempts for \toolname{}.

\paragraph{Validated Refactoring}
We count an instance as a validated refactoring if the approach detects the correct function, surfaces it among the top ten candidate functions, and produces a non-trivial refactoring that passes Stage~4 validation, meaning it passes the differential test and the project tests. A non-trivial refactoring must replace hand-written logic with a call to the target library rather than only adding an import or making cosmetic changes. The generated edit must overlap with the developer-changed lines recorded in the benchmark; an edit that refactors a completely separate region is not counted as a validated refactoring, even if the edit is otherwise useful. The validated refactoring must also pass on the held-out tests of the benchmark.

\subsubsection{LLM}
\label{sec:eval:models}
\toolname{} uses the same model, \texttt{gpt-5-mini}, for detector generation, refactoring generation, and differential-test generation.
We run Codex CLI version \emph{v0.139.0} with \texttt{gpt-5-mini}, default temperature, and a single run per benchmark instance.

\subsection{Results}
\label{sec:eval:results}

\subsubsection{RQ1: End-to-End Effectiveness}
\label{sec:eval:rq1}

\begin{table}
\centering
\caption{RQ1 end-to-end effectiveness of \toolname{}. \emph{Inst.}=instances; \emph{File Det.}=files detected; \emph{Func. Det.}=functions detected; \emph{Refac.}=proposed refactoring; \emph{PDT}=passes differential test (after up to three repairs); \emph{Val.}=validated (passes PDT and benchmark tests).}
\vspace{-0.7em}

\label{tab:rq1-library-breakdown}
\footnotesize
\setlength{\tabcolsep}{4pt}
\begin{tabular}{lrrrrrr}
\toprule
\textbf{Library} & \textbf{Inst.} & \textbf{File Det.} & \textbf{Func. Det.} & \textbf{Refac.} & \textbf{\shortstack[c]{Pass Diff\\Test}} & \textbf{Val.} \\
\midrule
\texttt{click}       & 6  & 6 & 6 & 5 & 5 & 5 \\
\texttt{collections} & 4  & 2 & 2 & 2 & 2 & 2 \\
\texttt{dataclasses} & 7  & 6 & 5 & 5 & 5 & 5 \\
\texttt{fastapi}     & 3  & 2 & 2 & 2 & 2 & 2 \\
\texttt{functools}   & 6  & 4 & 3 & 3 & 3 & 3 \\
\texttt{git}         & 7  & 7 & 5 & 5 & 5 & 1 \\
\texttt{httpx}       & 6  & 5 & 5 & 5 & 5 & 0 \\
\texttt{numpy}       & 9  & 4 & 1 & 1 & 1 & 1 \\
\texttt{pandas}      & 5  & 1 & 0 & 0 & 0 & 0 \\
\texttt{psutil}      & 3  & 2 & 1 & 0 & 0 & 0 \\
\texttt{pydantic}    & 4  & 2 & 2 & 1 & 1 & 1 \\
\texttt{pytest}      & 4  & 2 & 2 & 1 & 1 & 1 \\
\texttt{pytorch}     & 4  & 3 & 3 & 2 & 2 & 2 \\
\texttt{re}          & 4  & 3 & 3 & 3 & 3 & 3 \\
\texttt{requests}    & 9  & 8 & 5 & 3 & 3 & 3 \\
\texttt{scipy}       & 7  & 7 & 7 & 7 & 7 & 7 \\
\texttt{shutil}      & 7  & 6 & 1 & 1 & 1 & 1 \\
\texttt{yaml}        & 5  & 5 & 3 & 3 & 3 & 3 \\
\midrule
\textbf{Total}       & \textbf{100} & \textbf{75} & \textbf{56} & \textbf{49} & \textbf{49} & \textbf{40} \\
\bottomrule
\end{tabular}
\end{table}

We first evaluate whether \toolname{} can complete the full library-adoption refactoring task end to end: detect a real opportunity, localize it to a concrete function, generate a library-based rewrite, and validate that the rewrite preserves behavior. Table~\ref{tab:rq1-library-breakdown} summarizes the results with a per-library breakdown. Starting from 100 benchmark instances, \toolname{} detects the target file in 75 instances and surfaces the correct function among the top ten candidate functions for generation in 56 instances. Of these 56 functions, \toolname{} proposes a candidate refactoring for 49; in the seven remaining instances the generator concludes that no suitable target-library refactoring is available. All 49 proposed refactorings pass the differential test (with up to three repair attempts). Nine subsequently fail on held-out tests, leaving 40 validated refactorings.

Table~\ref{tab:rq1-library-breakdown} also shows that effectiveness is spread across many target libraries. \toolname{} produces validated refactorings for 15 of the 18 libraries in the benchmark and surfaces the reference function in 17 of 18 libraries. The strongest results occur for libraries where the benchmark refactorings expose both API-related vocabulary and recognizable implementation structure, such as \texttt{scipy}, \texttt{dataclasses}, \texttt{click}, \texttt{requests}, \texttt{yaml}, and \texttt{re}. For example, \toolname{} validates all seven \texttt{scipy} refactorings, all five \texttt{dataclasses} refactorings that reach generation, and five of six \texttt{click} instances.

The weaker cases reveal where the pipeline loses candidates. For \texttt{numpy} and \texttt{pandas}, \toolname{} reaches only four of nine \texttt{numpy} files and one of five \texttt{pandas} files, surfacing only one correct function for \texttt{numpy} and none for \texttt{pandas}. Manual inspection suggests that these libraries expose broad and heterogeneous API surfaces: opportunities may involve array construction, aggregation, indexing, reshaping, missing-value handling, type conversion, or numerical kernels, each with different vocabulary and implementation structure. Treating such libraries as single targets may therefore be too coarse; a more specialized pipeline could generate detectors for narrower API families or submodules of a target library. Other libraries fail mainly at ranking rather than initial detection. For example, \texttt{shutil} has six file-level detections but only one function-level detection, indicating that \toolname{} often reaches the right file but does not rank the target function highly enough.

To understand the 60 instances that do not reach a validated refactoring, we assign each one to the first pipeline stage it is lost. Table~\ref{tab:failure-modes} summarizes these mutually exclusive categories. \emph{No detector match} means no generated detector matched the target. \emph{Not in Top-10} means that at least one detector matched the project, but the reference function was not selected among the ten functions sent to generation; this includes both file-budget losses (target file not in the top five) and function-budget losses (target function not ranked high enough within the selected files). \emph{No refactoring} and \emph{Validation failure} cover instances where the reference function was surfaced but generation or validation did not succeed.

\begin{table}
\centering
\caption{Failure modes among the 60 instances that do not reach a validated refactoring.}
\vspace{-0.7em}
\label{tab:failure-modes}
\setlength{\tabcolsep}{1pt}
\begin{tabular}{lrl}
\toprule
\textbf{Failure mode} & \textbf{Count} & \textbf{Interpretation} \\
\midrule
No detector match        &  9 & No generated detector matched the target \\
Not in Top-10            & 35 & Target outside file or function budget \\
No refactoring           &  7 & LLM does not suggest a refactoring \\
Validation failure       &  9 & Generated refactoring fails benchmark tests \\
\bottomrule
\end{tabular}
\end{table}

\begin{figure}
\centering

\begin{beforebox}[title={BEFORE \textnormal{(\texttt{sktime/.../frequencies.py})}}]
\begin{lstlisting}[style=pythonpretty]
def _get_intervals_count_and_unit(freq: str):
    if freq is None:
        raise ValueError("frequency is missing")
\end{lstlisting}
\begin{delband}
\begin{lstlisting}[style=pythonpretty]
    m = re.match(r"(?P<count>\d*)(?P<unit>[a-zA-Z]+)$", freq)
    if not m:
        raise ValueError(f"pandas frequency {freq} not understood.")
    count, unit = m.groups()
    count = 1 if not count else int(count)
    return count, unit
\end{lstlisting}
\end{delband}
\end{beforebox}

\vspace{2pt}

\begin{afterbox}[title={AFTER \textnormal{(intended refactoring; not detected by \toolname{})}}]
\begin{lstlisting}[style=pythonpretty]
def _get_intervals_count_and_unit(freq: str):
    if freq is None:
        raise ValueError("frequency is missing")
\end{lstlisting}
\begin{addband}
\begin{lstlisting}[style=pythonpretty]
    offset = pandas.tseries.frequencies.to_offset(freq)
    return offset.n, offset.name
\end{lstlisting}
\end{addband}
\end{afterbox}

\vspace{2pt}
\scriptsize{(a) Detection miss: \texttt{sktime} opportunity for a rare \texttt{pandas} API.}

\vspace{6pt}

\begin{beforebox}[title={BEFORE \textnormal{(\texttt{Megatron-LM/.../clip\_grad.py})}}]
\begin{lstlisting}[style=pythonpretty]
def _clip_grad_norm(parameters, max_norm, norm_type=2):
\end{lstlisting}
\begin{delband}
\begin{lstlisting}[style=pythonpretty]
    is_not_tp_duplicate = param.tensor_model_parallel or \
        (mpu.get_tensor_model_parallel_rank() == 0)
    ...
    total_norm_cuda = torch.cuda.FloatTensor([float(total_norm)])
    torch.distributed.all_reduce(
    total_norm_cuda, op=torch.distributed.ReduceOp.SUM,
    group=mpu.get_model_parallel_group())
    total_norm = total_norm_cuda[0].item() ...
    
\end{lstlisting}
\end{delband}
\end{beforebox}

\vspace{2pt}

\begin{afterbox}[title={AFTER \textnormal{(attempted refactoring; validation rejected)}}]
\begin{lstlisting}[style=pythonpretty]
def _clip_grad_norm(parameters, max_norm, norm_type=2):
\end{lstlisting}
\begin{addband}
\begin{lstlisting}[style=pythonpretty]
    return torch.nn.utils.clip_grad_norm_(
        parameters, max_norm, norm_type=norm_type)
\end{lstlisting}
\end{addband}
\end{afterbox}

\vspace{2pt}
\scriptsize{(b) Refactoring failure: \texttt{Megatron-LM} behavior exceeds the PyTorch API.}

\caption{Representative failure examples.}
\label{fig:failure-examples}
\end{figure}

Detection misses arise when no generated detector matches the target file or function. These cases are most visible for large, heterogeneous libraries such as \texttt{numpy} and \texttt{pandas}, where opportunities span many API families and implementation styles. Figure~\ref{fig:failure-examples}a shows a \texttt{pandas} miss from \texttt{sktime}: the code performs generic regular-expression parsing and tuple unpacking, while the API used by the developer, \texttt{pandas.tseries.frequencies.to\_offset}, is a specialized frequency-parsing utility. The generated detectors therefore did not capture this specific part of the \texttt{pandas} API. This suggests that, for large libraries, detector generation may need to target narrower API families or submodules, for example, generating detectors specifically for \texttt{pandas.tseries} rather than for \texttt{pandas} as a whole.

A second source of failures is ranking quality. In these cases, \toolname{} may detect the relevant file or even match part of the reference function, but the function is not surfaced within the top candidates. This can happen when the target region receives only weak detector matches. It can also happen when other functions in the same file are plausible library-adoption opportunities and rank above the benchmark target. In a manual inspection of 10 higher-ranked non-reference candidates sampled from ranking failures, six were genuine library-adoption opportunities, just not the developer-refactored instances recorded in the benchmark.

The remaining failures happen during generation and validation. In seven cases, the model reports that no suitable library-based refactoring is available, often because the prompt does not identify the specific API needed for the rewrite. In another nine cases, the model generates a rewrite that passes the differential test but fails on project tests. Figure~\ref{fig:failure-examples}b shows a PyTorch example from Megatron-LM. The candidate function resembles \texttt{torch.nn.utils.clip\_grad\_norm\_}, but the hand-written implementation also includes tensor-model-parallel filtering and distributed all-reduce behavior. A naive replacement with the library call would remove this project-specific behavior, so \toolname{} rejects the generated edit instead of accepting a behavior-changing rewrite. The differential tests, which run on generated inputs, do not expose the parallel and distributed paths; the project tests do, causing the refactoring to be rejected.

\subsubsection{RQ2: Comparison to Baselines}
\label{sec:eval:rq2}

The primary comparison is candidate detection, because every baseline can report files or functions that may contain a library-adoption opportunity. For approaches that produce edits, we evaluate suggested refactorings using the same validation as for \toolname{}.

Table~\ref{tab:detection-results} summarizes the results. \toolname{} achieves the highest detection effectiveness, detecting 75 instances at the file level and 56 at the function level. The strongest baseline is Codex CLI, which detects 35 instances at the file level and 32 at the function level. The vanilla LLM baseline detects 32 files but only 7 functions, showing that direct prompting can identify some relevant files but struggles to narrow the search to the correct function. Semgrep+LLM reaches 30 file-level and 23 function-level detection, while linter-based tools detect only 11 and 8, respectively.

\begin{table}
\centering
\caption{Comparison against baselines on the 100-instance benchmark.}
\vspace{-0.7em}

\label{tab:detection-results}
\begin{tabular}{lccc}
\toprule
\textbf{Approach} & \textbf{File Det.} & \textbf{Func. Det.} & \textbf{Val.} \\
\midrule
Linters              & 11 &  8 & {--} \\
Vanilla LLM          & 32 &  7 & 4 \\
Semgrep+LLM          & 30 & 23 & {--} \\
Codex CLI            & 35 & 32 & 22 \\
\toolname{}          & \textbf{75} & \textbf{56} & \textbf{40} \\
\bottomrule
\end{tabular}
\end{table}

Rule-based tools are precise for the patterns they encode but cannot generalize to new ones. Semgrep+LLM attempts to close this gap by generating rules with an LLM, but LLM-written rules remain syntactically rigid and cannot generalize to the implementation variations present in real-world hand-written code; none of its 23 function-level detections translate to validated refactorings because the approach has no refactoring phase. The vanilla LLM baseline reveals a different bottleneck: file selection works because choosing likely files requires only coarse judgment over a project structure, but function localization collapses without a code-level search signal.

\subsubsection{RQ3: Efficiency and Cost}
\label{sec:eval:rq3}

To evaluate efficiency, we report scan time, LLM calls, and monetary cost. Scan time is the CPU cost of the scan stage alone; it differs from end-to-end wall-clock time, which is dominated by LLM-call latency across all four stages.  All experiments are on a 20-core virtual machine (31\,GiB RAM, Linux x86-64). For \toolname{}, we distinguish detector-generation cost from per-scan cost, because generated detectors can be reused across project scans.

We break down costs across three scopes: detector generation is charged once per project-library pair; scanning and ranking are charged per project scan and require no LLM calls; refactoring generation is charged per selected candidate function. Unless otherwise stated, the end-to-end per-instance cost includes detector generation, scanning, ranking, refactoring generation, differential-test generation, and any repair attempts for that benchmark instance.

Detector generation costs \$0.03 per project-library pair on average and takes about 80 seconds. Using those detectors, scanning a target project takes 1.3 seconds on average, with no LLM calls, while ranking and filtering complete in less than a second. As a result, after detector generation, \toolname{} can rescan a project in only a few seconds without spending additional LLM budget. Refactoring generation uses LLM calls only for selected candidate functions, keeping this stage inexpensive: \$0.01 per candidate function and 2.5 LLM calls per candidate function on average.

The strongest baseline, Codex CLI, spends about \$0.06 per instance while detecting fewer correct instances. It also pays this search cost every time it is run: reapplying Codex CLI to the same project requires a new agentic search, which can produce different candidate sets across runs making results harder to reproduce and coverage of a project less predictable. In contrast, \toolname{} turns LLM output into reusable detectors, so repeated scans are cheap and deterministic.

\subsubsection{RQ4: Ablation Study and Design Contributions}
\label{sec:eval:rq4}

We study which parts of \toolname{} contribute most to detection and end-to-end refactoring. The ablations follow the main design choices in the pipeline: the detector type used in Stage~1, the ranking budget used in Stage~3, and the repair loop used during validation.

\paragraph{Detector type}
We first ablate the two detector types generated in Stage~1: lexical detectors (regex-based detection) and structural detectors (AST-based matching). Table~\ref{tab:detector-ablation} reports the results. Lexical detectors alone find opportunities in 47 files and 31 functions. Structural detectors alone cover a similar number of files, 48, but slightly fewer functions, 30. Combining the two detector types substantially improves detection, reaching 75 file-level and 56 function-level detections. This shows that the two detector types are complementary: each family alone detects roughly 30 functions, while together they detect 56, indicating that lexical and structural cover different opportunities.

\begin{table}
\centering

\caption{Effect of detector type on candidate detection. Counts report files and functions detected out of 100 benchmark instances.}
\vspace{-0.7em}
\label{tab:detector-ablation}
\begin{tabular}{lcc}
\toprule
\textbf{Detector type} & \textbf{File Det.} & \textbf{Func. Det.} \\
\midrule
Lexical only & 47 & 31 \\
Structural only & 48 & 30 \\
Lexical + structural & \textbf{75} & \textbf{56} \\
\bottomrule
\end{tabular}
\end{table}

\paragraph{Ranking budget}
We next vary the number of files and functions retained before refactoring generation. The default configuration keeps the top five files and the top ten functions overall from those files.
Table~\ref{tab:ranking-budget} shows the recall-cost tradeoff. With only the top file, \toolname{} detects 43 functions, so a strict file cutoff loses several valid targets. Increasing the file budget from one to three recovers most of the lost recall, and increasing from three to five files adds two more functions.

\begin{table}
\centering
\caption{Effect of ranking budget on function-level detection.}
\vspace{-0.7em}
\label{tab:ranking-budget}
\begin{tabular}{lcc}
\toprule
\textbf{Ranking budget} & \textbf{Func. Det.} & \textbf{Gen. cost} \\
\midrule
Top 1 file,  top 10 func. & 43 & $1\times$ \\
Top 3 files, top 10 func. & 54 & $1\times$ \\
\textbf{Top 5 files, top 10 func.} & \textbf{56} & $1\times$ \\
Top 3 files, top 20 func. & 56 & $2\times$ \\
Top 5 files, top 20 func. & 58 & $2\times$ \\
\bottomrule
\end{tabular}
\end{table}

\begin{figure}
\centering
\includegraphics[width=0.9\columnwidth]{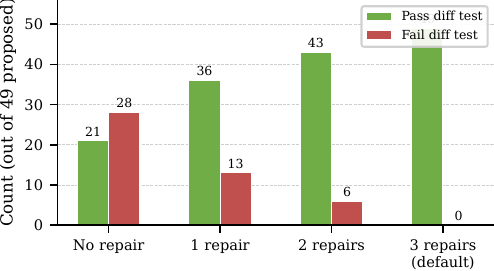}
\vspace{-1em}
\caption{Effect of repair budget on differential-test outcomes (out of 49 proposed). Green: pass the differential test; red: still fail it.}
\label{fig:repair-ablation}
\end{figure}

\paragraph{Repair loop}
Finally, we evaluate the repair loop used during refactoring validation. When the differential test fails or is inconclusive, \toolname{} feeds the failing input and observed outputs back to the LLM and asks it to repair the edit. This ablation asks whether validation feedback improves the final number of validated refactorings.
Figure~\ref{fig:repair-ablation} shows that repair feedback steadily converts failing proposals. With only the initial generation, 21 of 49 proposals pass the differential test and 28 still fail. Each repair round shrinks the failing group: after three repairs, all 49 pass the differential test. Of these, 40 are also validated by project tests; the remaining 9 pass the differential test but are rejected by project tests.

\subsubsection{RQ5: Open-Source Usefulness}
\label{sec:eval:rq5}

The fixed benchmark evaluates whether \toolname{} can recover historical developer refactorings, but this setting has two limitations. First, historical commits may appear in model training data, so benchmark recovery alone cannot rule out memorization effects. Second, a benchmark success does not show whether the generated refactorings are useful to maintainers of current projects. We therefore run \toolname{} on the latest versions of five benchmark projects and submit selected refactorings to the corresponding open-source maintainers.

We select projects whose latest versions still install and test successfully under our harness, and run \toolname{} with the corresponding target libraries from the benchmark. For each project, \toolname{} reports at most ten candidate functions. We manually inspect the final candidates and find that 67\% correspond to real library-adoption opportunities. To avoid overwhelming maintainers with refactoring pull requests, we submit one refactoring per repository, choosing a small validated edit that is easy to review.

Two of the five submitted pull requests have already been accepted by maintainers. Figure~\ref{fig:open-source-pr} shows one such accepted refactoring: a recursive Cartesian-product implementation replaced by a single \texttt{itertools.product} call, reducing 17 lines to 3. These results provide initial evidence that \toolname{} can surface actionable refactorings on current project versions, not only on mined historical commits.

\begin{figure}
\centering

\begin{beforebox}[title={BEFORE}]
\begin{lstlisting}[style=pythonpretty]
from typing import List, Any
...
def create_combinations(combinations: List[Any]) -> List[Any]:
\end{lstlisting}
\begin{delband}
\begin{lstlisting}[style=pythonpretty, escapechar=|]
    def __create_combinations(combinations_inner: List[Any]) -> List[Any]:
        ...  |\textcolor{diffrem}{\textbf{$\langle$+14 removed lines$\rangle$}}|
    return [combo for combo in arr if len(combo) == count]
\end{lstlisting}
\end{delband}
\end{beforebox}

\vspace{2pt}

\begin{afterbox}[title={AFTER \textnormal{(accepted pull request)}}]
\begin{addband}
\begin{lstlisting}[style=pythonpretty]
from itertools import product
\end{lstlisting}
\end{addband}
\begin{lstlisting}[style=pythonpretty]
...
def create_combinations(combinations: List[Any]) -> List[Any]:
\end{lstlisting}
\begin{addband}
\begin{lstlisting}[style=pythonpretty]
    if not combinations:
        return []
    return [list(combo) for combo in product(*combinations)]
\end{lstlisting}
\end{addband}
\end{afterbox}

\caption{Accepted pull request: recursive Cartesian-product enumeration replaced by \texttt{itertools.product} (17 lines to 3 lines).}
\label{fig:open-source-pr}
\end{figure}

\section{Limitations and Threats to Validity}
\label{sec:limitations}

\toolname{} handles function-level, behavior-preserving replacements. It does not address refactorings spanning files, requiring architectural changes, or modifying public interfaces. It rejects code that extends a library API with project-specific logic, since direct replacement would drop that behavior.
Without a generated detector match on the target file or function, an opportunity cannot reach ranking or generation. This is most visible for large, heterogeneous libraries such as \texttt{numpy} and \texttt{pandas}, with API families that use different vocabulary and shapes. Specialized API-family or submodule detectors could improve coverage.
We use the original implementation as the oracle and combine LLM-generated differential tests with project tests when available. However, tests cannot prove semantic equivalence. Validated refactorings may differ on untested inputs, while correct refactorings may be rejected because differential tests allow no divergence, even when benign.
\benchmark{} is mined from real developer refactorings, but reflects only committed opportunities and may omit other valid refactorings. Our implementation and evaluation focus on Python; the detector representations and validation workflow would need adaptation for other languages. Baseline results also depend on prompts, tool configurations, model versions, and search budgets; we mitigate this with comparable target-library hints and, for Codex CLI, the same model as \toolname{}.

\section{Related Work}
\label{sec:related}

\paragraph{Refactoring detection and mining}
Much work detects refactorings by mining repository histories. RefactoringMiner and its successors~\cite{tsantalis2018refactoringminer,tsantalis2022refactoringminer2} compare consecutive program versions and classify edits into Fowler-style categories~\cite{fowler2018refactoring}. Downstream studies use these tools to study refactoring frequency and motivation~\cite{silva2016whywerefactor,murphyhill2012howwerefactor}. These approaches mine existing history; \toolname{} instead finds and applies new library-adoption opportunities.

\paragraph{Custom-to-API replacement and API migration}
The closest prior work studies replacing custom implementations with API calls. RETIWA~\cite{tufano2022retiwa} and AKIRA~\cite{nyirongo2026akira} detect or recommend custom Java logic replaceable by an API, but do not generate or validate the resulting edits. \toolname{} differs in three ways: it targets library-adoption refactoring in Python projects, generates candidate edits, and accepts edits only after relevance filtering and execution-based validation.

LibraryMigration~\cite{teyton2012mining}, A3~\cite{lamothe2022a3}, and API-usage recommendation systems such as MAPO and DeepAPI~\cite{zhong2009mapo,gu2016deepapi} rely on existing API usages, migration examples, or usage patterns. Library adoption is harder because the replaced code may never reference the target library. \toolname{} therefore generates detectors for hand-written implementations and combines lexical and structural evidence with project context before invoking an LLM to rewrite selected candidates. Since API documentation improves LLM code generation for less-common libraries~\cite{chen2025apidocrag}, \toolname{} includes the target library and relevant API hints in the refactoring-generation prompt.

\paragraph{Static modernizers and linters}
Static modernizers and linters encode refactoring rules. Python tools such as pyupgrade~\cite{pyupgrade}, refurb~\cite{refurb}, Ruff~\cite{ruff}, Pylint~\cite{pylint}, and flake8 plugins~\cite{flake8} detect modernization and simplification opportunities with manually written rules; Zhang et al.~\cite{zhang2022pythonic} extend this to automatically rewrite non-idiomatic Python using AST transformations for predefined Pythonic idioms. The broader tradition of semantic patching, exemplified by Coccinelle~\cite{padioleau2008coccinelle} for Linux C code, showed that executable AST-level patterns can be applied systematically across large codebases. Large-scale experience with such analyses confirms that they deliver reliable, actionable results for covered patterns, but extending coverage to new domains requires substantial ongoing manual effort~\cite{sadowski2015tricorder,sadowski2018lessons}. \toolname{} targets this gap by generating detectors automatically from library metadata.

\paragraph{LLM agents for software engineering}
LLM agents support end-to-end software engineering tasks, including issue resolution in SWE-Agent and SWE-bench~\cite{yang2024sweagent,jimenez2024swebench}, project setup and test execution in ExecutionAgent~\cite{bouzenia2025executionagent}, and interactive coding with Aider and Claude Code~\cite{gauthier2024aider,anthropic2025claudecode}. These systems show that LLMs can navigate repositories and produce edits, but open-ended agentic search can be expensive and hard to control when scanning many projects or target libraries. \toolname{} uses LLMs in bounded roles: one generates reusable detectors offline, deterministic scanning applies them, and another is invoked only for top-ranked candidates.

\paragraph{Validating generated edits}
Generated code can be evaluated by static checks, LLM judges, or execution. LLM-as-judge methods~\cite{zheng2023judging} scale assessment, but judgment alone cannot establish behavior preservation. Execution-grounded benchmarks such as SWE-bench~\cite{jimenez2024swebench} and systems such as ExecutionAgent~\cite{bouzenia2025executionagent} instead emphasize running project tests. \toolname{} follows this view: LLM-generated differential tests pre-filter and guide repair, but final acceptance depends on relevance filtering and project tests.

\section{Conclusion}
\label{sec:conclusion}
This paper introduced \toolname{}, an automated approach for library-adoption refactoring in Python. On \benchmark{}, \toolname{} detects the correct file in 75 of 100 instances, surfaces the correct function in 56, and produces 40 validated refactorings, outperforming all baselines at roughly one-third the cost of the strongest agentic competitor. The failure analysis points to what remains difficult: large, heterogeneous libraries need finer-grained detector targeting, and I/O-heavy or distributed code resists differential validation. These results show that capturing LLM knowledge in reusable, executable detectors is more effective and cheaper than applying LLM reasoning through repeated project-wide queries.

\bibliographystyle{IEEEtran}
\bibliography{references}

\end{document}